\documentstyle[11pt,aasms4]{article}

\tighten
\slugcomment{Submitted for publication in 
{\it The Astrophysical Journal Letters}}

\begin{document}



\def\pmb#1{\setbox0=\hbox{#1}%
  \kern0.00em\copy0\kern-\wd0
  \kern0.03em\copy0\kern-\wd0
  \kern0.00em\raise.04em\copy0\kern-\wd0
  \kern0.03em\raise.04em\copy0\kern-\wd0\box0 }

\def\pp{\parshape=2 -0.25truein 6.75truein 0.5truein 6truein}

\def\ref #1;#2;#3;#4;#5{\par\pp #1 #2, #3, #4, #5}
\def\book #1;#2;#3{\par\pp #1 #2, #3}
\def\rep #1;#2;#3{\par\pp #1 #2, #3}

\def\undertext#1{$\underline{\smash{\hbox{#1}}}$}
\def\simlt{\lower.5ex\hbox{$\; \buildrel < \over \sim \;$}}
\def\simgt{\lower.5ex\hbox{$\; \buildrel > \over \sim \;$}}

\def\etal{{et~al.}}
\def\noi{\noindent}
\def\bs{\bigskip}
\def\ms{\medskip}
\def\ss{\smallskip}
\def\ob{\obeylines}
\def\l{\line}
\def\hrf{\hrulefill}
\def\hf{\hfil}
\def\q{\quad}
\def\qq{\qquad}
\renewcommand{\deg}{$^{\circ}$}
\newcommand{\uk}{$\mu$K}
\newcommand{\qrms}{$Q_{rms-PS}$}
\newcommand{\cdmr}{$COBE$-DMR}
\newcommand{\x}{$\otimes$}
\newcommand{\xrms}{$\otimes_{RMS}$}


\bs
\bs
\bs
\bs
\bs
\bs
\bs
\bs
\title{RMS Anisotropy 
in the $COBE$\altaffilmark{1}-DMR Four-Year Sky Maps}

\author{A.J. Banday\altaffilmark{2,3,4}, K.M. G\'orski\altaffilmark{2,5}, 
        C.L. Bennett\altaffilmark{6},
         G. Hinshaw\altaffilmark{2}, A. Kogut\altaffilmark{2},
         C. Lineweaver\altaffilmark{7}, \\
         G.F. Smoot\altaffilmark{8} \& L. Tenorio\altaffilmark{9}.
}

\noindent
\altaffiltext{1}{The National Aeronautics and Space Administration/Goddard 
Space Flight Center (NASA/GSFC) is responsible for the design, development, 
and operation of the {\it Cosmic Background Explorer (COBE)}. Scientific 
guidance is provided by the $COBE$ Science Working Group. GSFC is also 
responsible for the development of the analysis software and for the 
production of the mission data sets.}
\altaffiltext{2}{Hughes STX Corporation, LASP, Code 685, NASA/GSFC Greenbelt MD 20771.}
\altaffiltext{3}{e-mail: {\it banday@ceylon.gsfc.nasa.gov}}
\altaffiltext{4}{Max Plank Institut f\"ur Astrophysik, 85740 Garching bei M\"unchen, Germany.}
\altaffiltext{5}{on leave from Warsaw University Observatory, 
                 Aleje Ujazdowskie 4, 00-478 Warszawa, Poland.}
\altaffiltext{6}{NASA Goddard Space Flight Center, Code 685, 
                 Greenbelt MD 20771.}
\altaffiltext{7}{Observatoire de Strasbourg, Strasbourg, France.}
\altaffiltext{8}{LBL, SSL, \& CfPA, Bldg 50-25, University of California, 
                 Berkeley CA 94720.}
\altaffiltext{9}{Universidad Carlos III, Madrid, Spain.}

\bs
\bs
\bs

\begin{abstract}

The sky-RMS is the simplest model-independent characterization 
of a cosmological anisotropy signal.
The RMS temperature fluctuations determined from the
$COBE$-DMR four-year sky maps 
are frequency independent, consistent with the hypothesis that
they are cosmological in origin, with a typical
amplitude at 7\deg\ of $\sim\ 35\pm\, 2$ \uk\ and at 10\deg\ of
$\sim\ 29\pm\, 1$ \uk.
A joint analysis of the 7\deg\ and 10\deg\ \lq cross'-RMS derived from 
the data in both Galactic and Ecliptic coordinates
is used to determine the RMS quadrupole
normalization, $Q_{rms-PS}$, 
for a scale-invariant Harrison-Zel'dovich 
power law model. 
The difference in the inferred normalizations either including
or excluding the quadrupole is a consequence of the
sensitivity of the method to the Galaxy contaminated observed sky quadrupole.
Whilst there are variations depending on the data selection,
all results are consistent with 
an inferred \qrms\ normalization of $\sim\ 18\pm\, 2$ \uk.

\end{abstract}

\keywords{cosmic microwave background --- cosmology: observations}

\section{INTRODUCTION}

Previous analysis of results from the \cdmr\ experiment
(Smoot \etal\ 1992; Bennett \etal\ 1994)
has unambiguously demonstrated
the existence of cosmological anisotropy in the cosmic 
microwave background (CMB). The observed anisotropy is consistent with that
predicted by models of structure formation with power law initial
fluctuations of gaussian distributed amplitudes and random phases
(Bennett \etal\ 1996).

The sky-RMS signal is the simplest statistical measure of 
cosmological structure that can be computed from the sky maps.
Additionally, the observed sky-RMS on a given angular scale is, in 
principle, a convenient 
quantity to use in the normalization of cosmological models.
However, Wright \etal\ (1994) and subsequently Banday \etal\ (1994)
demonstrated at length that
it is essential to account for both instrument
specific details, such as the exact beam response function (rather than
using a gaussian approximation, for example), and data analysis 
specific details,
such as the subtraction of the best-fit monopole and dipole from
the maps, which perturbs the inferred normalization from that derived from 
standard analytic formulae. 

In this $Letter$, we compute the sky-RMS values from the complete
four-year \cdmr\ data set and use these values 
to infer the normalization of a Harrison-Zel'dovich, $n$~=~1, 
power law model.

\section{GALACTIC CUT}

Emission from the Galactic plane dominates the observed sky signal
at DMR frequencies but can not be modelled
with sufficient accuracy to enable its subtraction.
Instead, we excise those pixels near the Galactic
plane where the CMB anisotropy is necessarily contaminated by such emission.
For simplicity, previous DMR analyses (although see Hinshaw \etal; 1995),
attempted to minimize the contribution from Galactic emission
by retaining only those pixels for which 
the Galactic latitude $\mid {\rm b}\mid$ $>$ 20\deg.
However, visual inspection of the DMR maps
is suggestive of additional Galactic emission from regions
lying just outside this cut which are known to exhibit strong 
emission in the infrared part of the spectrum.
Cay\'on \& Smoot (1995) noted two regions associated with
known Galactic structure which showed non-CMB frequency
dependence and were therefore likely to be Galactic in origin.
Such features will inevitably contaminate the measured sky
anisotropy amplitude, and therefore we have attempted
to excise them from the four-year analysis.

Kogut \etal\ (1996a) have demonstrated a correlation between 
free-free and dust emission as traced by the $DIRBE$
long wavelength (100, 140 and 240 $\mu$m) sky maps
at high Galactic latitudes. Furthermore, Haslam \& Osborne (1987)
noted a strong association in the spatial distributions of the
emission from the inner Galactic plane as recorded by the $IRAS$ 
60 $\mu$m survey and radio surveys at 2.7 and 5 GHz. This correlation
with the thermal emission suggests that the dust responsible for the
diffuse IR emission in the plane is mixed with ionized gas
from extended low density HII regions. On the basis of these arguments,
we have used the $DIRBE$ 140 $\mu$m sky map (after a model of zodiacal 
emission has been removed -- Reach \etal, 1995) to trace the 
strong Galactic emission around the Galactic plane.
To maintain continuity with earlier treatments, we also remove
all pixels at latitudes ${\rm \mid b\mid\ < 20^\circ}$. 
We find that the additional regions to be excluded from our analysis
are the \lq flares of obscuration' noted by de Vaucouleurs (1956)
in Scorpius and Ophiuchus in the northern Galactic hemisphere
and Taurus and Orion in the south.
The new cut leaves 3881 surviving pixels in Galactic coordinates, 
and 3890 in Ecliptic\footnote{\small{A 
further extension of the cut in which we also
exclude all pixels at ${\rm \mid b\mid\ < 30^\circ}$ has also
been computed, leaving 3056 and 3039 pixels in Galactic and
Ecliptic coordinates respectively. Lists of the surviving pixels
for each cut are available from the Banday on request.}}.

\section{SKY SIGNAL ESTIMATION}

The \cdmr\ four-year sky maps are used in the present investigation.
Kogut \etal\ (1996b) have demonstrated that the corresponding 
systematic error contributions are 
small, and we ignore them in this analysis (the 68\% upper limit
at 7\deg\ is $\sim$ 1.4 \uk).
A discussion of the effect of residual high latitude Galactic emission 
is deferred until Section 5. However, we note that since synchrotron,
dust and free-free Galactic emission 
have spatial morphologies characterized by a steeply falling power spectrum
($\sim\ \ell^{-3}$ -- Kogut \etal\ 1996a), only structure
on large angular scales is compromised. As a simple initial test of
the extent to which Galactic emission affects our results, we 
perform all computations both including and excluding the observed
quadrupole.

The analysis proceeds as described for the two-year data
in Banday \etal\ (1994) where
the cross-RMS, \xrms, between two maps $a$ and $b$ is defined in their eqn (1).
The data without additional smoothing is referred
to as 7\deg\ smoothing, since the central lobe of the DMR beam is {\it
approximately} described by a 7\deg\ FWHM gaussian.
For determining the sky-RMS at 10\deg, the data surviving
the Galactic cut are smoothed by a 7\deg\ FWHM gaussian kernel
with uniform weighting. 

Table 1 shows the observed \xrms\ values determined from independent
sky maps at 31.5, 53 and 90 GHz constructed from the first two years and second
two years of \cdmr\ data. The results are consistent with the 
presence of sky signal in all four years of data.
We have computed a frequency dependence in the \xrms\
of the form $\nu^{\beta}$  between pairs of frequencies. Since the three DMR
frequencies span the expected minimum in the Galactic foreground emission
(see Fig. 2 in Bennett \etal\ 1992 and Fig. 3 in Kogut \etal\ 1996a) then an average
spectral index $\beta$ computed
over the three frequencies could appear flat even with
considerable contamination and erroneously be ascribed to a small
Galactic contribution. 
The lack of any frequency dependence in the 
\xrms\ (expressed in thermodynamic temperature units) 
without any correction for Galactic emissions is consistent 
with the dominant signal being cosmic in origin\footnote{\small{At 53 and 90 GHz,
note that this is also consistent
with a small and equivalent Galactic contribution 
at each frequency as suggested by the results in Kogut \etal\ 1996a; 1996c}}.


\section{LIKELIHOOD ANALYSIS}

In this {\it Letter} we restrict our attention to the inference of
the \qrms\ normalization for the $n$ = 1 power spectrum model.
The analytical form of the probability distribution
of the \cdmr\ data analysis specific \xrms\ statistic is 
unmanagable. A Monte Carlo approach was adopted to generate
the \xrms\ distributions for a grid of \qrms\ values
(with 5000 simulations used for each value of \qrms).
Each simulation
generates maps of the sky temperature distribution by combining a
realization of Harrison-Zel'dovich sky anisotropy filtered through the \cdmr\
beam (Wright \etal\ 1994).
The expected variance of the spherical harmonic
coefficients of the CMB temperature in this case is
given by $< a_{lm}^2 >\ \propto\ 1/[\ell\, (\ell\, +\,1)]$ (Abbott \& Wise 1984).
Noise realizations are based on the appropriate values
of the RMS per observation 
and observation patterns of the specific DMR channels.
The effects of noise correlations (Lineweaver \etal\ 1994, G\'orski \etal\ 1996a)
are not significant for this work
and hence are not included in the simulations.
The \qrms-dependent statistical means, variances
and covariances of the 7\deg\ and 10\deg\ \xrms\
are derived from these Monte Carlo simulations, and used to
construct the gaussian approximation to the probability
distribution of the \xrms.
This, together with the measured \xrms\ values, defines the
likelihood function $\cal L$(\qrms).

We consider \xrms\ values computed between the A and B channels
at each of the three frequencies, between the 53 and 90 GHz (A+B)/2 or sum (S)
maps, and between A and B maps made by coadding the corresponding
sides at the three frequencies. The inverse-noise-variance weights for the
31.5, 53 and 90 GHz maps in thermodynamic temperature are then 
(0.090,0.718,0.192) and (0.062,0.588,0.350) for the A and B
channels respectively.

Table 2 summarizes the four-year \xrms\ values 
and the inferred maximum likelihood values of \qrms. 
The small observed differences between the Galactic and
Ecliptic values are expected since there is some noise rebinning
depending on the pixelization scheme.
The normalizations determined at each frequency or from 
combinations of frequency agree at the $1\ \sigma$ level. 
These values are completely 
consistent with previous analyses based on either the first year
or first two years of data, lying between these earlier normalizations,
and are in excellent agreement with values derived by more sensitive techniques
using the four-year data 
(see the accompanying papers G\'orski \etal\ 1996b; Hinshaw \etal\ 1996;
Wright \etal\ 1996). 

Since any likelihood technique is only unbiased asymptotically,
we have tested the \xrms\ technique described above with
an additional set of simulations with noise levels appropriate
to the 53S\x 90S combination.
There is evidence for a small, statistically significant bias (uncorrected in Table 2) 
to a lower \qrms. Over a range of input
normalizations from 15 to 20 \uk, the bias is typically $\sim$ 0.2 \uk.
The bias is sensitive to the signal-to-noise ratio, which is
why Banday \etal\ 1994 saw no evidence for this effect in their
analysis of the two-year data

An additional check on the bias correction can be determined from a
one-dimensional version of the (\qrms, $n$) matching technique
described in Wright \etal\ (1996). The cosmic part of a simulated map
can be considered as a function of \qrms. Thus, the \xrms\
value of a simulation and the resultant inferred quadrupole 
normalization generated by processing through the standard maximum likelihood ($ML$)
machinery, ($Q_{ML,\, sim}$), is itself a function of \qrms. For any given sky realization,
we can choose a normalization amplitude
-- $Q_{match,\, sim}$ -- which results in the same
inferred normalization as the real data. For 
2500 simulations matched to the observed sky signal, the mean
$Q_{match,\, sim}$ is a good estimate of \qrms. For the 53S\x 90S Galactic
combination, we find \qrms\ = $18.4^{+1.7}_{-2.1}$ \uk\ ($\ell\ \geq\ 3$).

\section{DISCUSSION}

A comparison of results including or excluding the quadrupole
shows evidence of a systematic shift to a lower
\qrms\ normalization when the quadrupole is included.
The effect should not be considered as surprising, however.
Kogut \etal\ (1996c) have demonstrated that the cosmic quadrupole
is counter-aligned to the Galactic quadrupole, and thus the
quadrupole as observed in sky-maps uncorrected for 
Galactic contamination is not representative of the 
CMB quadrupole. 
Whilst the shift is within the range of uncertainties expected from
noise and cosmic variance, it undoubtedly reflects the
partial cancellation of Galactic and cosmological quadrupoles.
Furthermore, since the RMS estimator (at a given resolution)
compresses all of the available information in the data to one number,
it is very sensitive to the $\ell$ = 2 mode. A simple calculation
(not accounting for noise or sky-coverage) for an $n$ = 1
spectrum shows that $\sim$ 23\% of the sky variance 
over the spectral range $\ell\ \in$ [2,40]
is contributed by the quadrupole. 

The inferred value for \qrms\ using the \xrms\ technique is therefore
sensitive to the choice of quadrupole amplitude in the sky-maps.
We have estimated this bias using a set of simulations 
where the power spectrum normalization 
has been varied over the range 15 - 20 \uk,
whilst the realization specific quadrupole amplitude
was constrained to lie in the range [4, 28] \uk\ (Kogut \etal\ 1996c).
The sample data sets processed through the maximum likelihood machinery
yielded biases in the range -1.4 to -2.8 \uk,
sufficiently large to correct \qrms\ and reconcile the normalizations.

If we attempt a more specific correction, the foreground emission can be modelled
using the $DIRBE$-140 $\mu$m sky map as a template for free-free and
dust emission, and a full-sky radio survey at 408 MHz (Haslam \etal\ 1981)
to trace the synchrotron contribution (see Kogut \etal\ 1996b for discussion).
After subtraction of the appropriate Galactic contribution,
the normalization amplitudes including or excluding the quadrupole 
are in better agreement (Fig. 1): correcting the DMR maps
for Galactic emission increases the CMB quadrupole. 
The Galactic correction is itself subject to considerable uncertainty,
and when taken into account $all$ \qrms\ values are consistent with 
$\sim$ 18 \uk.


\section{CONCLUSIONS}

In summary, we have used the \xrms\ statistic derived from the
\cdmr\ four-year sky maps to infer a normalization
\qrms\ $\sim$ 18 \uk\ (with a typical significance of $\sim\ 9\, \sigma$)
for the amplitude of primordial inhomogeneity
in the context of a Harrison-Zel'dovich $n$ = 1 model. 
The sensitivity of the \xrms\ method to the Galaxy contaminated 
observed sky quadrupole 
affords a reasonable explanation for the difference in inferred \qrms\
amplitudes when either including or excluding the quadrupole. 
Normalizations inferred from both Galactic and Ecliptic coordinate
data sets at all frequencies agree at better than the $1\ \sigma$ level. 
The \cdmr\ four-year sky maps are reliable data sets for the investigation 
of cosmological models.

\ms

We acknowledge the efforts of those contributing to the $COBE$ DMR.
$COBE$ is supported by the Office of Space Sciences of NASA Headquarters.
AJB thanks Mark Hindmarsh for hospitality at the University of Sussex.
\clearpage

\begin{table}[h]
\noindent Table 1: Observed \xrms\ in thermodynamic temperature 
between maps of the first two and second two years of \cdmr\ data
at 31.5, 53 and 90 GHz. The errors are determined from 
noise simulations for a fixed CMB sky realization. The frequency
dependence of the data, expressed as \xrms($\nu$) $\propto\ \nu^{\beta}$
is indicated between pairs of frequencies.
No statistically significant frequency dependence is observed.
\begin{center}

\vspace{2mm}

\begin{tabular}{lcccccccc} \hline\hline
 & \multicolumn{4}{c}{Galactic} & \multicolumn{4}{c}{Ecliptic} \\ \cline{2-9}
Freq & \multicolumn{2}{c}{$\ell\ \geq\ 2$} &
\multicolumn{2}{c}{$\ell\ \geq\ 3$} & \multicolumn{2}{c}{$\ell\ \geq\ 2$} &
\multicolumn{2}{c}{$\ell\ \geq\ 3$} \\ \cline{2-9}
 & 7\deg\ (\uk) & 10\deg\ (\uk) & 7\deg\ (\uk) & 10\deg\ (\uk)
 & 7\deg\ (\uk) & 10\deg\ (\uk)  & 7\deg\ (\uk) & 10\deg\ (\uk) \\ \hline
31 & $42.5^{+26.9}_{-45.8}$ & $27.8^{+7.1}_{-5.0}$ & $38.5^{+29.6}_{-40.4}$ & $21.4^{+7.5}_{-5.2}$ 
   & $68.0^{+25.1}_{-46.6}$ & $31.1^{+6.8}_{-5.4}$ & $65.9^{+28.2}_{-42.5}$ & $26.3^{+7.7}_{-5.2}$ \\
53 & $31.9^{+2.6}_{-2.1}$ & $28.2^{+1.1}_{-1.3}$ & $31.6^{+3.1}_{-2.2}$ & $27.9^{+1.2}_{-1.3}$ 
   & $28.8^{+2.7}_{-2.4}$ & $28.7^{+1.2}_{-1.3}$ & $28.6^{+2.5}_{-2.3}$ & $28.5^{+1.3}_{-1.3}$ \\
90 & $30.2^{+6.5}_{-5.4}$ & $29.3^{+2.4}_{-2.1}$ & $29.9^{+6.8}_{-4.7}$ & $29.0^{+2.2}_{-2.0}$ 
   & $14.6^{+7.0}_{-4.8}$ & $29.5^{+2.4}_{-2.5}$ & $14.2^{+6.3}_{-5.6}$ & $29.3^{+2.1}_{-2.1}$ \\
\hline\hline
$\beta_{53:31}$ & $-0.6^{+1.2}_{-2.1}$ & $0.0^{+0.5}_{-0.4}$ & $-0.4^{+1.5}_{-2.0}$ & $0.5^{+0.7}_{-0.5}$ 
                & $-1.7^{+0.7}_{-1.3}$ & $-0.2^{+0.4}_{-0.3}$ & $-1.6^{+0.8}_{-1.2}$ & $0.2^{+0.6}_{-0.4}$ \\
$\beta_{90:53}$ & $-0.1^{+0.4}_{-0.4}$ & $0.1^{+0.2}_{-0.2}$ & $-0.1^{+0.5}_{-0.3}$ & $0.1^{+0.2}_{-0.2}$ 
                & $-1.3^{+0.9}_{-0.7}$ & $0.1^{+0.2}_{-0.2}$ & 
$-1.3^{+0.9}_{-0.8}$ & $0.1^{+0.2}_{-0.2}$ \\ 
$\beta_{90:31}$ & $-0.3^{+0.6}_{-1.0}$ & $0.1^{+0.3}_{-0.2}$ & $-0.2^{+0.8}_
{-1.0}$ & $0.3^{+0.3}_{-0.2}$ & $-1.5^{+0.6}_{-0.7}$ & $-0.1^{+0.2}_{-0.2}$ &
$-1.5^{+0.6}_{-0.7}$ & $0.1^{+0.3}_{-0.2}$ \\ \hline
\end{tabular}

\end{center}

\end{table}

\clearpage

\begin{table}[h]
\noindent Table 2: \xrms\ combinations used as input to the 
likelihood ($\cal L$) analysis and the inferred normalization
parameter, \qrms, for an assumed $n$ = 1 spectrum.
\begin{center}

\vspace{2mm}

\begin{tabular}{lrcccc} \hline\hline
\multicolumn{2}{c}{} & \multicolumn{2}{c}{Galactic} & \multicolumn{2}{c}{Ecliptic} \\ \cline{3-6}
combination & \multicolumn{1}{c}{(\uk)} & $\ell\ \geq\ 2$ & $\ell\ \geq\ 3$ & $\ell\ \geq\ 2$ 
& $\ell\ \geq\ 3$ \\ \hline
          &  7\deg\ RMS & $47.3^{+25.6}_{-39.5}$ & $43.7^{+25.0}_{-39.4}$ 
                              & $64.7^{+23.8}_{-45.0}$ & $62.4^{+24.1}_{-42.1}$ \\
31A\x 31B & 10\deg\ RMS & $29.6^{+5.9}_{-4.8}$ & $23.9^{+6.6}_{-5.0}$ 
                              & $32.4^{+5.8}_{-5.1}$ & $27.6^{+6.6}_{-4.7}$ \\
        &        \qrms\ & $16.6^{+4.1}_{-4.9}$ & $16.6^{+4.4}_{-6.6}$ 
                              & $17.9^{+3.7}_{-4.5}$ & $18.6^{+4.1}_{-5.8}$ \\
\multicolumn{6}{c}{} \\
          &  7\deg\ RMS & $36.2^{+3.2}_{-2.4}$ & $35.9^{+2.7}_{-2.3}$ 
                              & $33.2^{+3.6}_{-2.6}$ & $32.8^{+2.5}_{-2.3}$ \\
53A\x 53B & 10\deg\ RMS & $29.0^{+1.4}_{-1.2}$ & $28.6^{+1.3}_{-1.3}$ 
                              & $29.4^{+1.2}_{-1.3}$ & $29.0^{+1.3}_{-1.1}$ \\
        &        \qrms\ & $15.3^{+2.3}_{-1.7}$ & $18.2^{+2.2}_{-1.9}$ 
                              & $16.0^{+2.3}_{-1.9}$ & $18.7^{+2.1}_{-1.9}$ \\
\multicolumn{6}{c}{} \\
          &  7\deg\ RMS & $25.0^{+8.2}_{-4.9}$ & $24.4^{+7.4}_{-4.7}$
                              & $15.7^{+9.1}_{-5.8}$ & $14.7^{+6.7}_{-5.5}$ \\
90A\x 90B & 10\deg\ RMS & $27.0^{+2.5}_{-2.3}$ & $26.4^{+2.2}_{-2.0}$
                              & $29.0^{+2.8}_{-2.0}$ & $28.6^{+2.1}_{-2.5}$ \\
        &        \qrms\ & $14.8^{+2.3}_{-2.1}$ & $17.1^{+2.5}_{-2.2}$
                              & $15.1^{+1.9}_{-2.0}$ & $17.4^{+2.0}_{-1.8}$ \\
\multicolumn{6}{c}{} \\
          &  7\deg\ RMS & $32.5^{+2.5}_{-2.3}$ & $32.1^{+2.3}_{-2.0}$ 
                              & $31.3^{+2.5}_{-2.1}$ & $30.9^{+2.2}_{-1.9}$ \\
53S\x 90S & 10\deg\ RMS & $28.5^{+1.2}_{-1.1}$ & $28.0^{+1.2}_{-0.9}$
                              & $29.4^{+1.3}_{-1.2}$ & $29.0^{+1.1}_{-1.0}$ \\
        &        \qrms\ & $15.5^{+2.2}_{-1.8}$ & $18.1^{+2.1}_{-1.8}$
                              & $15.9^{+2.1}_{-1.7}$ & $18.3^{+2.0}_{-1.6}$ \\
\multicolumn{6}{c}{} \\
          &  7\deg\ RMS & $35.1^{+1.8}_{-1.8}$ & $34.7^{+1.6}_{-2.0}$ 
                              & $31.3^{+2.1}_{-1.7}$ & $30.8^{+1.9}_{-1.7}$ \\
coadded (A\x B) & 10\deg\ RMS & $28.9^{+0.9}_{-1.1}$ & $28.4^{+0.9}_{-0.9}$
                              & $29.5^{+0.9}_{-0.9}$ & $29.0^{+0.9}_{-1.0}$ \\
        &        \qrms\ & $15.5^{+2.4}_{-1.8}$ & $18.4^{+2.3}_{-1.9}$ 
                              & $15.8^{+1.9}_{-1.7}$ & $18.2^{+1.8}_{-1.6}$ \\
\hline

\end{tabular}

\end{center}

\end{table}

\clearpage

\begin{center}
FIGURE CAPTIONS
\end{center}

\vspace{7mm}

\noindent Fig 1. Likelihood curves for \qrms\ derived using the 53S\x 90S 
\xrms\ combination at 7\deg\ and 10\deg\ assuming $n$ = 1.
Thick lines represent the analysis based
on the Galactic data, thin lines correspond to Ecliptic data analysis.
The dashed curves are for the quadrupole included ($\ell\ \geq\ 2$) case, 
solid curves exclude the quadrupole ($\ell\ \geq\ 3$).
A consistent shift to lower \qrms\ is evident in the quadrupole included
fits.
The horizontal error bar shows the range of central \qrms\ values
determined from the quadrupole included data after plausible Galaxy corrections
have been applied. All \qrms\ values inferred from the \cdmr\ data
irrespective of data selection are consistent with a normalization
of $\sim$ 18 \uk\ (solid circle).

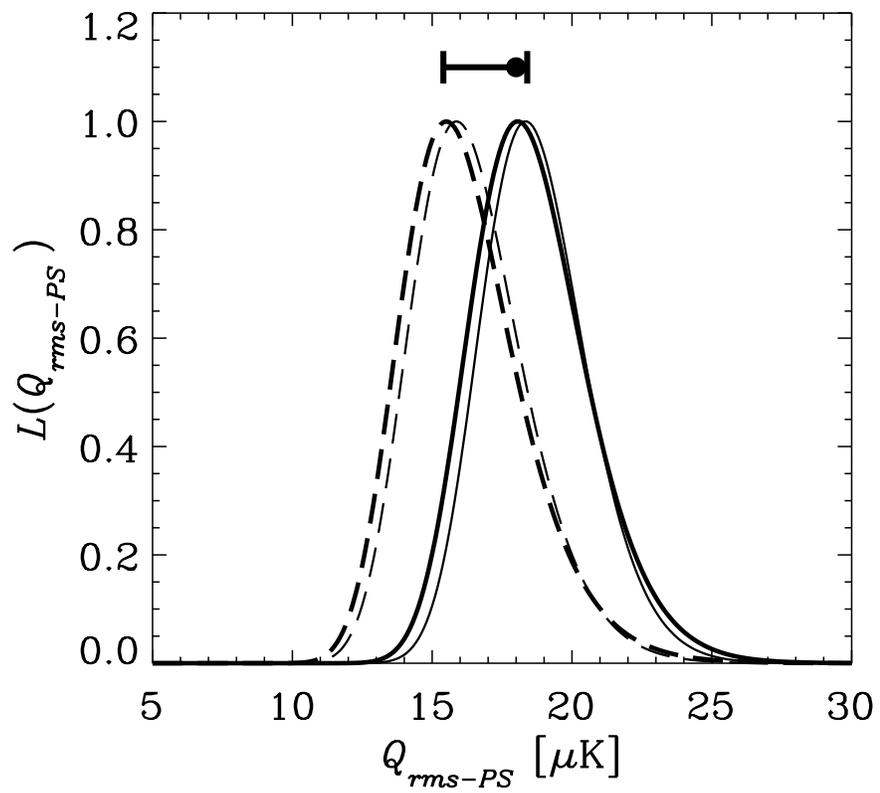
\end{document}